\documentclass[aps,pra,showpacs,twocolumn]{revtex4-1}
\usepackage{amsfonts}
\usepackage{amsmath}
\usepackage{amssymb}
\usepackage{graphicx}%
\usepackage{color}
\usepackage{epsfig}
\providecommand{\U}[1]{\protect\rule{.1in}{.1in}}

\begin{document}
\preprint{ }
\title{Electron momentum distributions and photoelectron spectra of atoms driven by intense spatially inhomogeneous field}
\author{M. F. Ciappina$^{1,2}$}
\author{J. A. P\'erez-Hern\'andez$^{3}$}
\author{T. Shaaran$^{1}$}
\author{L. Roso $^{3}$}
\author{M. Lewenstein$^{1,4}$}
\affiliation{$^{1}$ICFO-Institut de Ci\`encies Fot\`oniques, Mediterranean Technology Park, 08860 Castelldefels (Barcelona), Spain}
\affiliation{$^{2}$Department of Physics, Auburn University, Auburn, Alabama 36849, USA}
\affiliation{$^{3}$Centro de L\'aseres Pulsados, CLPU, Parque Cient\'{\i}fico, 37185 Villamayor, Salamanca, Spain}
\affiliation{$^{4}$ICREA-Instituci\'o Catalana de Recerca i Estudis Avan\c{c}ats, Lluis Companys 23, 08010 Barcelona, Spain}

\keywords{above threshold ionization; nanostructures; plasmonics; metal nanoparticles}
\pacs{42.65.Ky,78.67.Bf, 32.80.Rm}
\begin{abstract}

We use three dimensional time-dependent Schr\"odinger equation (3D--TDSE) to calculate angular electron momentum distributions and photoelectron spectra of atoms driven by spatially inhomogeneous fields. An example for such inhomogeneous fields is the locally enhanced field induced by resonant plasmons, appearing at surfaces of  metallic nanoparticles, nanotips and gold bow-tie shape nanostructures. Our studies show that the inhomogeneity of the laser electric field plays an important role in the above threshold ionization process in the tunneling regime, causing  significant modifications to the electron momentum distributions and photoelectron spectra, while its effects in the multiphoton regime appear to be negligible. Indeed, through tunneling ATI process, one can obtain higher energy electrons as well as high degree of asymmetry in the momentum space map. In this study we consider near infrared laser fields with intensities in the mid- $10^{14}$ W/cm$^{2}$ range and we use linear approximation to describe their spatial dependence. We show  that in this case it is possible to drive electrons with energies in the near-keV regime. Furthermore, we study how the carrier envelope phase influences the emission of ATI photoelectrons for few-cycle pulses. Our quantum mechanical calculations are fully supported by their classical counterparts.
\end{abstract}
\maketitle

\section{Introduction}

The process known as above-threshold ionization (ATI), in which an atom or molecule absorbs more photons than the minimum number required to single ionize it, has been a subject of intensive studies during the last decades (see e.g. \cite{milosevic_rev} and references therein). The first experimental realization was made at the end of the seventies~\cite{Agostini1979} and since then there has been a truly amazing progress in understanding of the non-perturbative nature of ATI. The recent advances in laser technology made possible to routinely generate laser pulses with few cycles of duration, which allows control of the atomic and molecular processes in their natural time scales, i.e. in the range of (sub)-femto to attoseconds. In addition, these short laser sources find an extensive range of applications in basic science, such as controlling molecular motions and chemical reactions ~\cite{schnurer2000,vdHoff2009}. Furthermore, the few-cycles pulses provide the fundamental pillar in the generation of high-order harmonics and the creation of isolated extreme ultraviolet (XUV) pulses~\cite{ferrari2010,schultze2007}.

The appearance of COLTRIMS experiments (see e.g.~\cite{coltrims}
and references therein) offered an unprecedented  possibility of
performing stringent tests on the different theoretical
approaches. On one side, this is because the imaging of the
vectorial momentum distributions of the reaction fragments are
easily accessible, while on the other they are particularly
sensitive to various details of the theory. COLTRIMS were
primarily developed for the study of few-body dynamics induced by
particle impact, i.e. electrons and ions, but the extension to
scrutinize and tackle laser-induced processes was natural (see
e.g~\cite{artem1,maharjan,schuricke}). Among the features which
were theoretically analyzed was the complex emission pattern
present in the two-dimensional momentum plane, parallel and
perpendicular to the laser polarization axis, of the laser-ionized
electron distributions near threshold~\cite{arbo1}. It was also
investigated  how these patterns evolve as the laser-matter
process change from the multiphoton to the tunneling
regimes~\cite{schuricke}.

The main difference between a few-cycle pulse and a multicycle one
is the strong dependence of the laser electric field on the
so-called Carrier Envelope Phase
(CEP)~\cite{Wittmann2009,kling2008}. The electric field in a
few-cycle pulse can be characterized by its duration and by the
CEP. The influence of CEP has been experimentally observed in
high-harmonic generation (HHG)~\cite{nisoli2003}, the emission
direction of electrons from atoms~\cite{paulus2001}, and in the
yield of nonsequential double ionization~\cite{liu2004}. In order
to have a better control of the system on an attosecond temporal
scale it is, therefore, important to find reliable and direct
schemes to measure the absolute phase of few-cycle pulses.

The investigation of ATI generated by few-cycle driving laser
pulses plays a key role in the CEP characterization due to the
sensitivity of the energy and angle-resolved photoelectron spectra
to the value of the laser electric field absolute
phase~\cite{paulus_cleo,sayler}. Consequently, the behavior of the
laser-ionized electrons renders the ATI phenomenon a very valuable
tool for laser pulse characterization. To determine the CEP of a
few-cycle laser pulse, it is essential to record the difference
between the yield of electrons ionized for different emission
angles. Through this technique one can analysis the so-called
backward-forward asymmetry in order to obtain the absolute
CEP~\cite{paulus}. Furthermore, it appears that the high energy
region of the photoelectron spectra is most sensitive to the
absolute CEP and consequently electrons with large kinetic energy
are needed in order to describe
it~\cite{milosevic_rev,paulus2003}.

Recent experiments using a combination of plasmonic nanostructures
and rare gases have demonstrated that the harmonic cutoff of the
gases could be extended further than in conventional situations by
using the field, locally enhanced due to the coupling of a laser
pulse with a nanosystems~\cite{kim,Kimnew}. In such nanosystems,
due to the strong confinement of the plasmonics spots and the
distortion of the electric field by the surface plasmons, the
locally enhanced field is not spatially homogeneous in the region,
where the electron dynamics take place. One should note, however,
that the outcome of the experiments of Ref.~\cite{kim}, in which a
combination of gold bow-tie nanostructure and argon gas for
generation HHG was used, has been recently under intense scrutiny
\cite{sivis,Kimreply}. In addition, recently, instead of atoms or
molecules in gas phase solid state nanostructures have been
employed as a target to study the photoelectron emission by few
intense laser pulses~\cite{kling,peternature}. This laser driven
phenomenon,  called Above Threshold Photoemission (ATP), has
received special attention due to the novelty of the involved
physics and potential applications. In ATP process, the emitted
electrons have energy far beyond the usual cutoff for noble gases
(see
e.g.~\cite{peterprl2006,peterprl2010,peternature,peterjpbreview,jensdombi,ropers}).
Furthermore, the photoelectrons emitted from these nanosources are
sensitive to the CEP,  and consequently this fact plays an
important role in the angle and energy resolved photoelectron
spectra~\cite{apolonski,dombi,kling,peternature}.

From theoretical point of view, the fundamental assumption behind
strong field phenomena, that the laser field is spatially
homogeneous in the region where the electron dynamics takes
place~\cite{keitel,krausz}, is not any more valid for the locally
enhanced plasmonic field. Indeed, in such system the driven
electric field, and consequently the Lorentz force the electron
feels, will also depend on position. Up to now, there have been
very few studies investigating the strong field phenomena in such
spatially inhomogeneous
fields~\cite{husakou,ciappi2012,yavuz,ciappi_opt,ciappiati1d,tahirsfa,tahirJMO}.
All of these studies have demonstrated that the spatial dependency
of the field strongly modifies the laser-driven phenomena that
appear in such circumstances.

For homogeneous driving field, up to know, different numerical and
analytical approaches have been employed to calculate the ATI (see
e.g.~\cite{milosevic_rev,schafer1993,telnov2009,bauer2006,Blaga2009,Quan2009}
and references therein). In this article, we extend the studies of
our previous paper~\cite{ciappiati1d} by applying the numerical
solution of the time-dependent Schr\"odinger Equation (TDSE) in
three dimensions to calculate the angular electron momentum
distributions and photoelectron spectra of ATI driven by spatially
inhomogeneous fields, considering both tunneling and multiphoton
regimes. The spatial dependence of the field is considered to be
linear. We mainly focus on studying ATI of hydrogen atoms, but our
scheme, within the single active electron approximation, can be
directly applied to any complex atom. We demonstrate how the
inhomogeneity of the field modifies both the two electron momentum
distributions and the photoelectron spectra; we examine also the
influence of the CEP parameter. Furthermore, our quantum
mechanical results are compared with classical calculations of the
kinetic energy of the electron.

This article is organized as follows. In the next section, we
present our theoretical approach to model ATI produced by
spatially nonhomogeneous fields, with main emphasis on the
extraction of the electron angular momentum distributions starting
from the TDSE outcomes. Subsequently, in Sec.~III, we apply our
method to compute the electron momentum distributions and
energy-resolved photoelectron spectra of hydrogen atom using
few-cycle laser pulses for both homogeneous and inhomogeneous
fields, considering tunneling and multiphoton regimes.
Furthermore, we solve the classical equations of motion of an
electron in an oscillating inhomogeneous electric field to support
our quantum mechanical method. Finally, in Sec.~IV, we conclude
our contributions with a short summary and outlook.

\section{Theoretical approach}

To study the properties of the ATI phenomenon driven by spatial
nonhomogeneous fields, we solve the three dimensional Time
Dependent Schr\"odinger Equation (3D-TDSE) in the length gauge.
The electron momentum distribution and energy-resolved
photoelectron spectra of an atom are calculated from the time
propagated electronic wave function.

Our calculations are based on spherical harmonics expansion,
$Y_{l}^{m}$, considering only the $m = 0$ terms due to the
cylindrical symmetry of the problem. The Crank-Nicholson method,
which is based on a splitting of the time-evolution operator that
preserves the norm of the wave function, is used as the numerical
technique. We consider the field to be linearly polarized along
the $z$ axis and the variation of the electric field is linear
with respect to the position. As a result, the coupling
$V_{l}(\mathbf{r},t)$ between the atomic system and the
electromagnetic radiation reads
\begin{equation}
\label{vlaser} V_{l}(\mathbf{r},t)=\int ^\mathbf{r}
d\mathbf{r'}\cdot\mathbf{E}(\mathbf{r'},t)=E_0z(1+\beta
z)f(t)\sin(\omega t+\phi),
\end{equation}
where $E_0$,  $\omega$ and  $\phi$ are the laser electric field
amplitude, the central frequency and characterizes the carrier
envelope phase (CEP), respectively. The parameter $\beta$ defines
the `strength'  of the inhomogeneity and has units of inverse
length (see also~\cite{husakou,yavuz,ciappi2012}). For modeling
the short laser pulses in Eq.~(\ref{vlaser}), we use a sin-squared
envelope $f(t)$ of the form
\begin{equation}
f(t)=\sin^{2}\left(\frac{\omega t}{2 n_p}\right),
\end{equation}
where $n_p$ is the total number of optical cycles. As a result,
the total duration of the laser pulse will be $T_p=n_p \tau$ where
$\tau=2\pi/\omega$ is the laser period. We also assume that before
switch on of the laser ($t=-\infty$) the target atom (hydrogen) is
in its ground state ($1s$), whose analytic form can be found in
standard textbooks. Within the single active electron
approximation, however, our numerical scheme is tunable to treat
any complex atom by choosing the adequate effective (Hartree-Fock)
potential, and finding the ground state by the means of numerical
diagonalization.

\begin{figure}[ht]
\resizebox{3in}{!}{\includegraphics[angle=270]{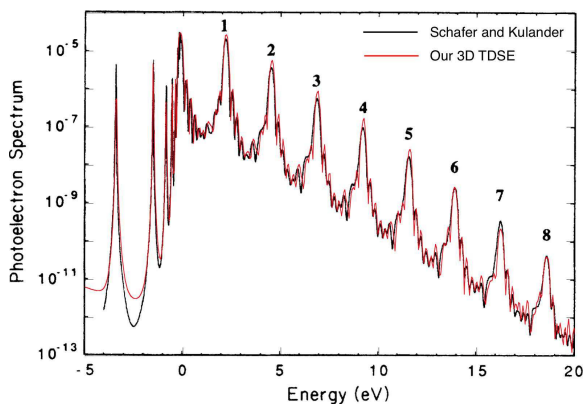}}
\caption{(Color online) Photoelectron spectrum resulting from our 3D TDSE simulations (in red) and superimposed (in black) with the ATI results calculated by Schafer and Kulander in Ref.~\cite{schaferwop1}. The laser wavelength is $\lambda=532$ nm and the intensity is $I=2\times10^{13}$ W/cm$^2$ (see Fig.1 in~\cite{schaferwop1} for more details. The superimposed plot has been extracted from Fig. 1 of this cited reference.}
\label{Fig1_comp}
\end{figure}

The ATI spectrum is calculated using the time dependent wave function method developed by Schafer and Kulander (see~\cite{schaferwop1} for more details). As a preliminary test, for ensuring the consistence of our numerical simulations, we have checked out our calculations with the results previously obtained in Ref.~\cite{schaferwop1}. The comparison confirms the high degree of accuracy of our calculations as shown in Fig.~\ref{Fig1_comp}.

For calculating the energy-resolved photoelectron spectra $P(E)$ and two-dimensional electron distributions $\mathcal{H}(p,\theta)$ we use the window function approach developed by Schafer~\cite{schaferwop1,schaferwop}.
This tool has been widely used, both to calculate angle-resolved and
energy-resolved photoelectron spectra~\cite{schaferwop2}, and it represents a
step forward with respect to the usual projection methods.

\section{Results}

In this section, we calculate both energy-resolved photoelectron
spectra $P(E)$ and two-dimensional electron momentum distributions in order to investigate
the influence of the inhomogeneities of the field and the sensitivity of these two measurable quantities to the different laser parameters, especially to the carrier envelope phase (CEP). The investigations are carried out for both the tunneling regime, for which the Keldysh parameters is $\gamma\lesssim1$ ($\gamma=\sqrt{I_p/2 U_p}$, where $U_p=I/4\omega^2$ is the ponderomotive energy and $I_p$ the ionization potential), and multiphoton regime, for which the Keldysh parameters is $\gamma>>1$. Furthermore, we confirm how in the tunneling regime the CEP, joint with the spatial nonhomogeneities, modify in a particular way both the energy-resolved photoelectron spectra and the two-dimensional electron momentum distributions as we have shown in our previous contribution~\cite{ciappiati1d}.  On the other hand, we show that in the multiphoton regime ($\gamma>>1$), the spatial nonhomogeneous character of the laser electric field hardly affects the analyzed quantities. We also want to point out, however, that the frontier between the tunnel and multiphoton regimes appears to be a controversial and diffuse issue~\cite{reiss2007,reiss2010}.

\subsection{Tunneling regime}

We commence by investigating the tunneling regime. For this case, we employ a four-cycle (total duration 10 fs) sin-squared laser pulse with wavelength $\lambda = 800$ nm and two different intensities, namely $I=1.140\times 10^{14}$ W/cm$^2$ and $I=5.0544\times 10^{14}$ W/cm$^2$. These two intensities give values for the laser electric field of $E_0=0.057$ a.u. and $E_0=0.12$ a.u., respectively.  For all the cases we chose four different values for the parameter that
characterizes the inhomogeneity strength, namely, $\beta=0$ (homogeneous case), 0.002, 0.003 and 0.005.  In addition we also vary the carrier envelope phase $\phi$ in Eq.~(\ref{vlaser}), taking $\phi=0$, $\phi=\pi/2$, $\phi=\pi$ and $\phi=3\pi/2$. For all the above mentioned cases, we calculate the the energy-resolved photoelectron spectra. The results are shown in Figures 2 and 3 for $I=1.140\times 10^{14}$ W/cm$^2$ ($\gamma=1$) and $I=5.0544\times 10^{14}$ W/cm$^2$ ($\gamma=0.475$), respectively.

\begin{figure}[ht]
\resizebox{0.5\textwidth}{!}{\includegraphics[angle=270]{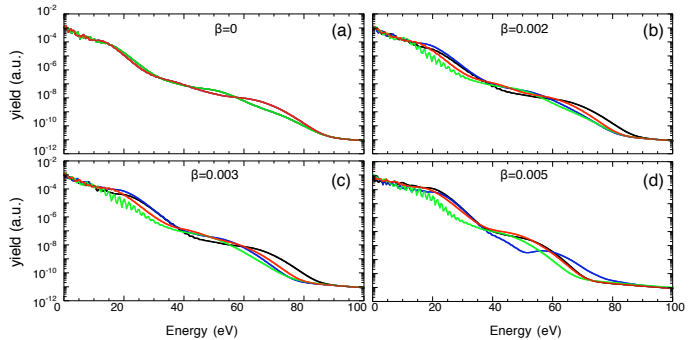}}
\caption{(Color online) Energy-resolved photoelectron spectra $P(E)$ calculated
using the 3D-TDSE for an hydrogen atom ($I_p = -0.5$ a.u.). The laser
parameters are $I = 1.140 \times 10^{14}$ W/cm$^2$ ($E_0=0.057$ a.u.) and $\lambda = 800$ nm. We have used
a sin-squared shaped pulse with a total duration of four optical cycles
(10 fs).  (a) $\beta = 0$
(homogeneous case), (b) $\beta = 0.002$, (c) $\beta = 0.003$ and (d) $\beta = 0.005$. In all the panels, black line: $\phi=0$; blue line $\phi=\pi/2$; green line: $\phi=\pi$ and red line: $\phi=3\pi/2$.}
\label{Fig2}
\end{figure}

\begin{figure}[ht]
\resizebox{0.5\textwidth}{!}{\includegraphics[angle=270]{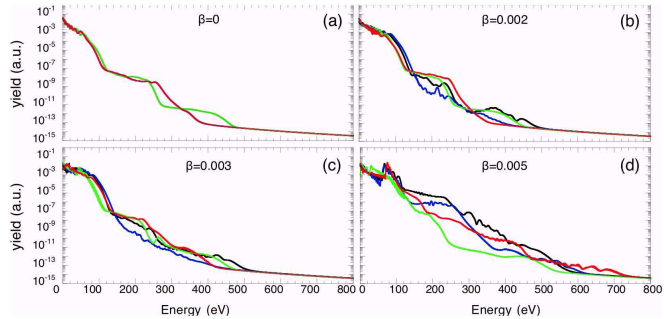}}
\caption{(Color online) Idem Fig. 2 but for a laser intensity $I = 5.404 \times 10^{14}$ W/cm$^2$ ($E_0=0.12$ a.u.).}
\label{Fig3}
\end{figure}

For the homogeneous case, the spectra exhibits the usual
distinct behavior, namely, the $2U_p$ cutoff ($\approx 13.6$ eV in Fig. 2(a) and  $\approx 57.4$ eV in Fig. 3(a)) and the $10U_p$ cutoff ($\approx 68$ eV in Fig. ~2(a) and $\approx 300$ eV in Fig.~3(a)). The former cutoff corresponds to those electrons that, once ionized, never return to the atomic
core (the so-called \textit{direct} electrons), while the latter one corresponds to the electrons that,
once ionized, return to the core and elastically rescatter (the so-called \textit{rescattered} electrons). Classically, it is
a well established arguments that the maximum
kinetic energies $E_k$ of the direct and the rescattered electrons
are $E^d_{max} = 2U_p$ and $E^r_{max}= 10U_p$, respectively (see below for more details). In a quantum
mechanical approach, however, it is possible to find electrons
with energies beyond the $10Up$ cutoff, although their yield
drops several orders of magnitude depending strongly on the atomic species studied~\cite{milosevic_rev}.

\begin{figure}[ht]
\resizebox{0.5\textwidth}{!}{\includegraphics[angle=270]{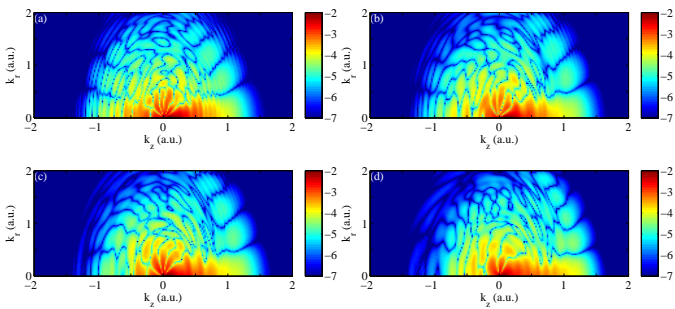}}
\caption{(Color online) Two-dimensional electron momentum
distributions (logarithmic scale) in cylindrical coordinates ($k_z,k_r$)
using the exact 3D-TDSE calculation for an hydrogen atom.
The laser parameters are $I = 1.140 \times 10^{14}$ W/cm$^2$ ($E_0=0.057$ a.u.) and $\lambda = 800$ nm. We have used
a sin-squared shaped pulse with a total duration of four optical cycles
(10 fs) with $\phi=0$. (a) $\beta = 0$
(homogeneous case), (b) $\beta = 0.002$, (c) $\beta = 0.003$ and (d) $\beta = 0.005$.}
\label{Fig4}
\end{figure}

Experimentally speaking, both the direct and rescattered electrons contribute to the energy-resolved photoelectron
spectra. It means for tackling this problem both physical mechanisms should be included in theoretical model. In that sense, the 3D-TDSE, which
can be considered as an exact approach to the ATI problem for atoms in the single active electron approximation, is the adequate tool to predict the $P(E)$ in the whole range of electron energies. On the other hand, the most energetic electrons, i.e., those
with kinetic energies $E_k \gg 2U_p$, are commonly used to characterize the
CEP of few-cycle pulses. Consequently, a correct description of the electron rescattering mechanism is needed.

\begin{figure}[ht]
\resizebox{0.5\textwidth}{!}{\includegraphics[angle=270]{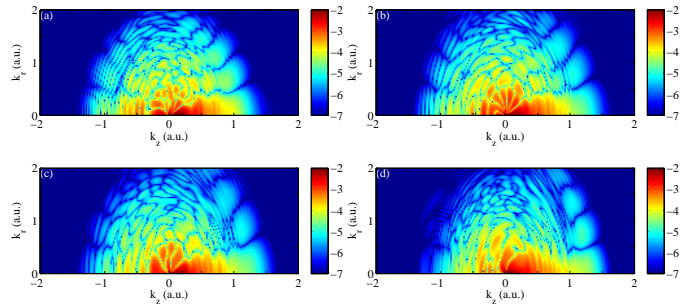}}
\caption{(Color online) Idem Fig. 4 but for $\phi=\pi/2$}
\label{Fig5}
\end{figure}

For the spatially nonhomogeneous cases,
the positions of the direct and the rescattered
electron cutoffs are extended towards larger
energies. For the rescattered electrons, this extension is very
noticeable. In fact for $\beta = 0.005$ with $E_0=0.12$ a.u., it reaches values close to $\approx700$ eV
[Fig. 3(d)] against the $\approx 300$ eV shown by the homogeneous case. Another new feature present for all the nonhomogeneous cases is the strong sensitivity of the $P(E)$ to the carrier envelope phase (CEP). This behavior can be clearly noticed by comparing the panels (a) of Figs. 2 and 3 (i.e. the homogeneous case) with the rest of the plots. It is clear that for the homogeneous case only two curves are present, due to the fact that the $P(E)$ for $\phi=0$ and ($\phi=\pi/2$) are identical to $\phi=\pi$ and ($\phi=3\pi/2$), respectively. On the other hand, for all the nonhomogeneous cases it is possible to clearly distinguish the 4 cases, i.e. $\phi=0$, $\phi=\pi/2$, $\phi=\pi$ and $\phi=3\pi/2$. Indeed, this particular characteristic of the $P(E)$ for nonhomogeneous fields could make them a new and better CEP characterization tool.

It should be noted, however, that other well-known and established CEP characterization tools,
such as, for instance, the forward-backward asymmetry or two-dimensional electron momentum distributions should complement the $P(E)$ measurements~\cite{milosevic_rev}. Furthermore, the utilization of nonhomogeneous fields would open the avenue for the production of high-energy electrons, reaching the keV regime, if a reliable control of the spatial and temporal shape of the laser electric field is attained. Investigations in such direction has already started (see e.g.~\cite{peterjpbreview} and references therein).

\begin{figure}[ht]
\resizebox{0.5\textwidth}{!}{\includegraphics[angle=270]{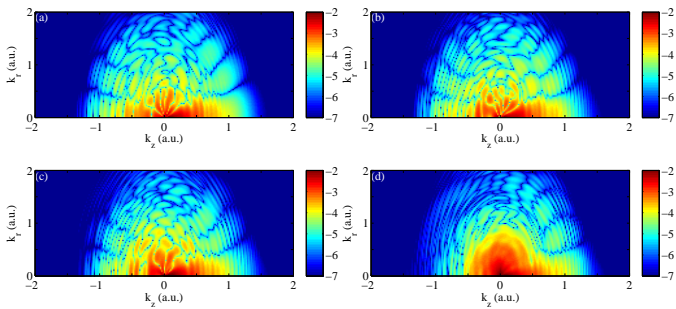}}
\caption{(Color online) Idem Fig. 4 but for $\phi=\pi$}
\label{Fig6}
\end{figure}

A deep analysis of the electron distributions for atomic ionization
produced by an external laser field can be performed
in terms of the two-dimensional electron momentum distribution.
The exact solution of the three dimensional Schr\"odinger equation (3D-TDSE) provide
us with an excellent tool to analyze in detail how the two competing fields, namely
the laser electric field and the Coulomb potential, modify the electron wavepacket of the released electron.
In Figs. 4-7 we calculate two-dimensional electron momentum distribution for a laser field with an intensity of $I=1.140\times10^{14}$ W/cm$^{2}$ ($E_0=0.057$ a.u), $\lambda=800$ nm and different values of the the $\beta$ parameter: panel (a) $\beta=0$ (homogeneous case); panel (b) $\beta=0.002$; panel (c) $\beta=0.003$ and panel (d) $\beta=0.005$. We employ a few-cycle laser pulse with 4 total cycles (10 fs) and various values of the carrier envelope phase (CEP) parameter $\phi$. Figs. 4-7 show the cases with  $\phi=0$, $\phi=\pi/2$, $\phi=\pi$ and $\phi=3\pi/2$, respectively.

\begin{figure}[ht]
\resizebox{0.5\textwidth}{!}{\includegraphics[angle=270]{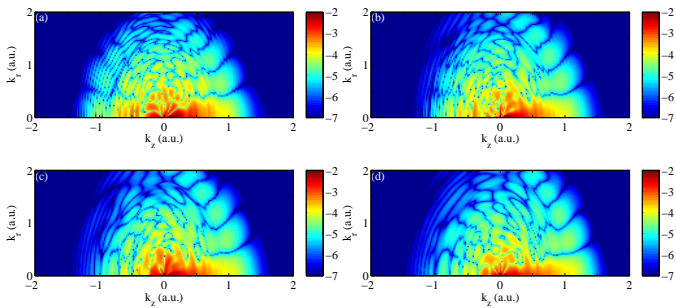}}
\caption{(Color online) Idem Fig. 4 but for $\phi=3\pi/2$}
\label{Fig7}
\end{figure}

Here, we concentrate our analysis on the low energy region of the
distributions in order to study how the inhomogeneities of the
laser electric field affect the angular electron yield. This
region shows the usual \textit{bouquet}-type structures
(see~\cite{arbo1} for details) with noticeable modifications for
the nonhomogeneous cases.

\begin{figure}[ht]
\resizebox{0.45\textwidth}{!}{\includegraphics[angle=270]{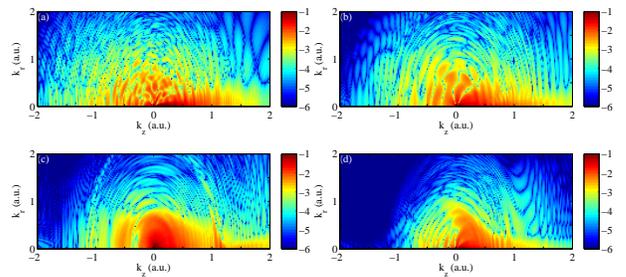}}
\caption{(Color online) Two-dimensional electron momentum
distributions (logarithmic scale) in cylindrical coordinates ($k_z,k_r$)
using the exact 3D-TDSE calculation for an hydrogen atom.
The laser parameters are $I = 5.0544 \times 10^{14}$ W/cm$^2$ ($E_0=0.12$ a.u.) and $\lambda = 800$ nm. We have used
a sin-squared shaped pulse with a total duration of four optical cycles
(10 fs) with $\phi=0$. (a) $\beta = 0$
(homogeneous case), (b) $\beta = 0.002$, (c) $\beta = 0.003$ and (d) $\beta = 0.005$.}
\label{Fig8}
\end{figure}

Furthermore, the low-energy electrons appear to be strongly
influenced by the spatial inhomogeneity of the laser electric
field (see panels (b)-(d) of Figs. 4-7). We also can observe how
the \textit{bouquet} structures present in the homogeneous case
\textit{disappear} for particular values of $\phi$ (see e.g.
Fig.6(d)).

In order to complete our investigations, we calculate
two-dimensional electron momentum distributions by increasing the
laser field intensity to $I=5.0544\times10^{14}$ W/cm$^{2}$
($E_0=0.12$ a.u). The results are depicted in Figs. 8-10 for
$\phi=0$, $\phi=\pi/2$, $\phi=\pi$ and $\phi=3\pi/2$,
respectively. Here, panel (a), (b), (c) and (d) represent the
cases with $\beta=0$ (homogeneous case), $\beta=0.002$,
$\beta=0.003$ and $\beta=0.005$, respectively.

\begin{figure}[ht]
\resizebox{0.45\textwidth}{!}{\includegraphics[angle=270]{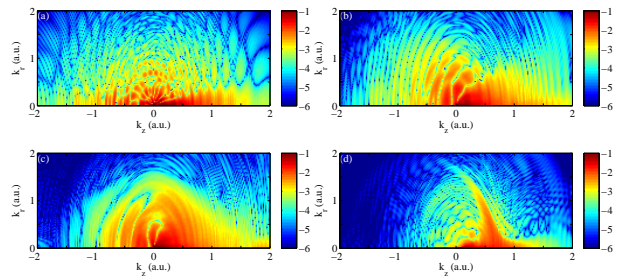}}
\caption{(Color online) Idem Fig. 8 but for $\phi=\pi/2$}
\label{Fig9}
\end{figure}

\begin{figure}[ht]
\resizebox{0.45\textwidth}{!}{\includegraphics[angle=270]{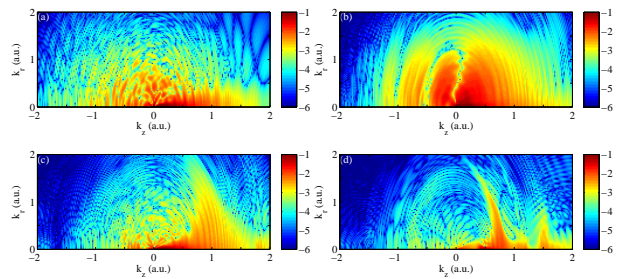}}
\caption{(Color online) Idem Fig. 8 but for $\phi=\pi$}
\label{Fig10}
\end{figure}

Following the trend observed in the previous studied case, we see
strong modifications produced by the spatial inhomogeneities in
both the angular and low-energy structures. Here, the modification
seems to be even more pronounced. For instance, the yield for
electrons with $k_z<-1$ a.u. for $\phi=0$ (Fig. 8(a)) and
$\phi=\pi/2$ (Fig. 9(a)) drops several orders of magnitude. The
significant drop of the yield of the electron emission between
$k_z<-1$ a.u and $k_z>0$ a.u. opens a new approach to characterize
the CEP.

We now employ a classical model in order to explain and characterize the extension of the energy-resolved photoelectron spectra. According to the simple-man's model~\cite{corkum} the physical mechanism behind ATI process will be as follows: at a given time, that we call ionization time $t_i$, an atomic electron is released or born at the position $z=0$ with zero
velocity, i.e., $\dot{z}(t_i)=0$. This electron now moves only under
the influence of the oscillating laser electric field (this model neglects the
Coulomb interaction with the remaining ion) and will reach
the detector either directly or through the process known as rescattering. By using the classical equation of motion of an electron
moving in an oscillating electric field, it is possible to
calculate the maximum energy of the electron for both the direct
and rescattered processes.

\begin{figure}[ht]
\resizebox{0.45\textwidth}{!}{\includegraphics[angle=270]{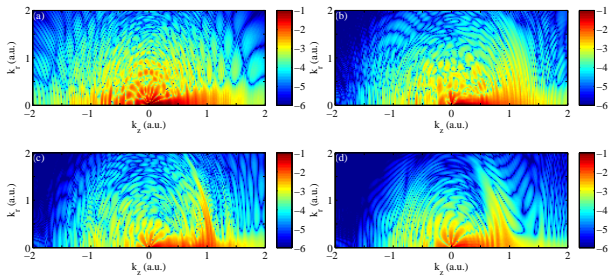}}
\caption{(Color online) Idem Fig. 8 but for $\phi=3\pi/2$}
\label{Fig11}
\end{figure}

\begin{figure}[ht]
\resizebox{0.4\textwidth}{!}{\includegraphics[angle=0]{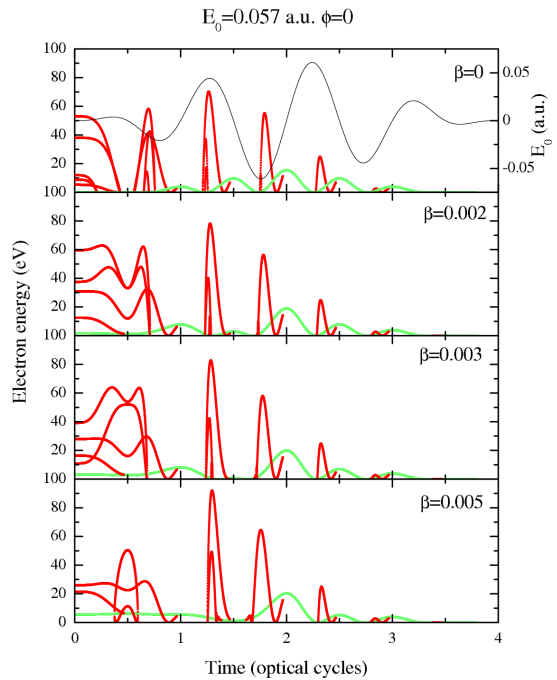}}
\caption{(Color online) Numerical solutions of the Newton equation [Eq.~(\ref{newton})] plotted in terms of the direct and rescattered electron kinetic energy, $E_{k}^{d}$  and $E_{k}^{r}$, respectively. The laser parameters are $I=1.140\times10^{14}$ W/cm$^{2}$ ($E_0=0.057$ a.u), $\lambda=800$ nm and $\phi=0$. We employ a few-cycle laser pulse with 4 total cycles (10 fs). Different panels correspond to various values of $\beta$ (see labels). Green filled circles: \textit{direct} electrons; red filled circles: \textit{rescattered} electrons.}
\label{Fig12}
\end{figure}

The Newton equation of motion for
the atomic electron in the laser electric field can be written, using the functional form of Eq.~(\ref{vlaser}), as
follows
\begin{eqnarray}
\label{newton}
\ddot{z}(t)&=&-\nabla_z V_{l}(\mathbf{r},t)\\
&=&E(t)[1+2\beta z(t)]
\end{eqnarray}
where we have collected the time-dependent part of the electric
field in $E(t)$, i.e., $E(t) = E_0 f (t) \sin(\omega t + \phi)$.

\begin{figure}[ht]
\resizebox{0.4\textwidth}{!}{\includegraphics{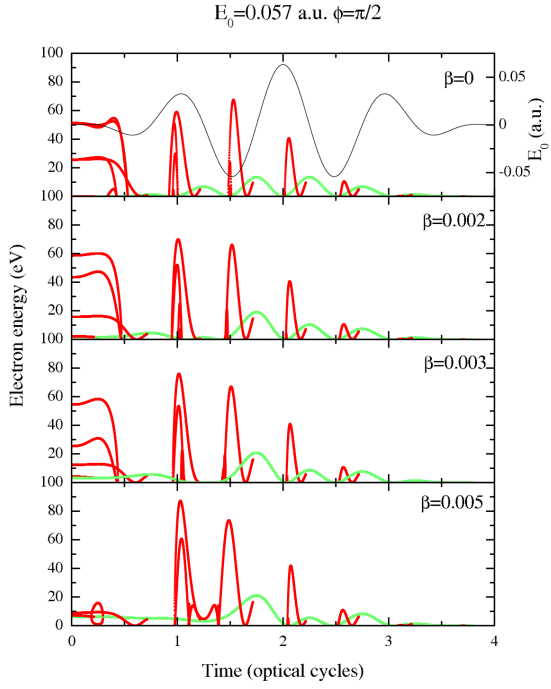}}
\caption{(Color online) Idem Fig. 12 for $\phi=\pi/2$.}
\label{Fig13}
\end{figure}

In the limit where $\beta=0$
in Eq.~(\ref{newton}), we recover the homogeneous case. For the direct
laser ionization, the kinetic energy of an electron released or born
at time $t_i$ is
\begin{equation}
\label{direct}
E^{d}_k=\frac{\left[ \dot{z}(t_{i})-\dot{z}(t_{f})\right]^{2}}{2},
\end{equation}
where $t_{f}$ is the end time of the laser pulse. For the rescattered
laser-ionized electron, in which the electron returns to the core at a time $t_{r}$ and
reverses its direction, the kinetic energy of the electron yields
\begin{equation}
\label{rescattered}
E^{r}_k=\frac{\left[ \dot{z}(t_{i})+\dot{z}(t_{f})-2\dot{z}(t_{r})\right] ^{2}}{2}.
\end{equation}

For homogeneous fields, Equations (\ref{direct}) and (\ref{rescattered}) become the usual expressions $E_k^{d}=\frac{\left[ A(t_{i})-A(t_{f})\right] ^{2}}{2}$ and  $E_{k}^r=\frac{\left[ A(t_{i})+A(t_{f})-2A(t_{r})\right] ^{2}}{2}$, with $A(t)$ being the laser vector potential $A(t)=-\int^{t} E(t')dt'$, respectively. For the case with $\beta=0$, it
can be shown that the maximum value for $E^{d}_k$ is $2U_{p}$ while for $E^{r}_k$ it is $10U_{p}$~\cite{milosevic_rev}. These two values appear as cutoffs in the energy resolved photoelectron spectrum as can be observed in panels (a) of Figs.~2 and 3.

\begin{figure}[ht]
\resizebox{0.4\textwidth}{!}{\includegraphics{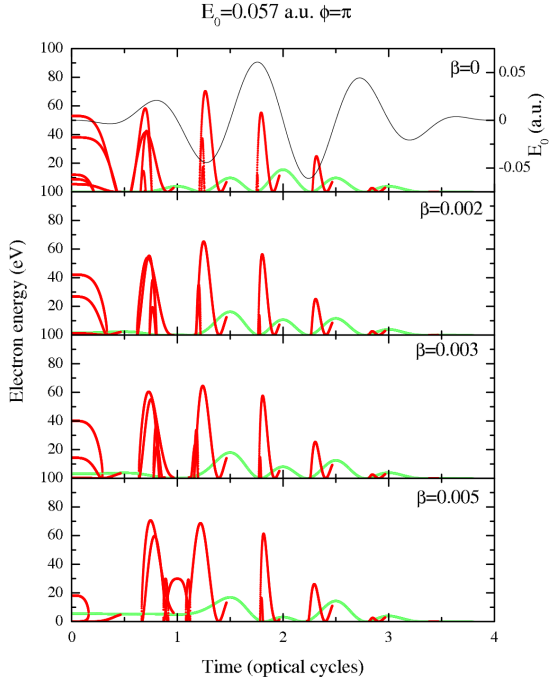}}
\caption{(Color online) Idem Fig. 12 for $\phi=\pi$.}
\label{Fig14}
\end{figure}

In Figs.~12-19, we present the numerical solutions of Eq.~(\ref{newton}), which is plotted in terms of the kinetic energy (in eV) of the direct and rescattered electrons.

\begin{figure}[ht]
\resizebox{0.4\textwidth}{!}{\includegraphics{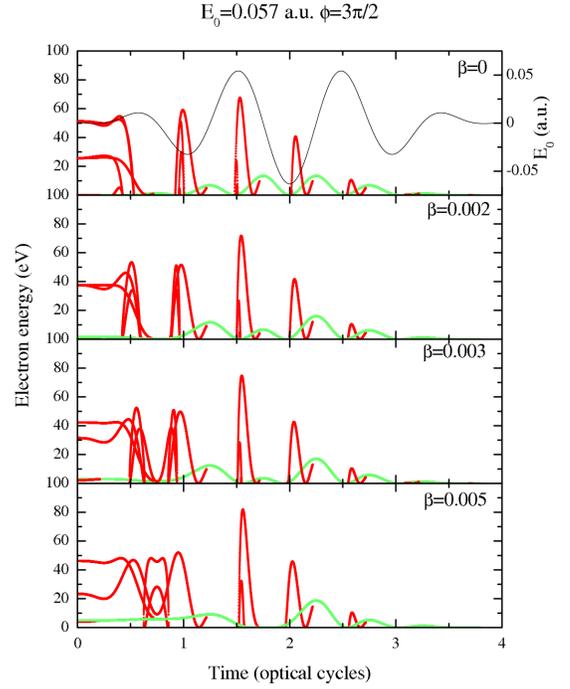}}
\caption{(Color online) Idem Fig. 12 for $\phi=3\pi/2$.}
\label{Fig15}
\end{figure}

Figures 12-15 are for a laser intensity of $I=1.140\times 10^{14}$ W/cm$^{2}$ ($E_0=0.057$ a.u.) meanwhile in Figs.~16-19 the laser intensity is $I=5.404\times 10^{14}$ W/cm$^{2}$ ($E_0=0.12$ a.u.).  Figures 12 (16), 13 (17), 14 (18) and 15 (19) are for $\phi=0$, $\phi=\pi/2$, $\phi=\pi$, $\phi=3\pi/2$, respectively and for different values of the $\beta$ parameter ($\beta=0$ (homogeneous case), $\beta=0.002$, $\beta=0.003$ and $\beta=0.005$, from top to bottom).

\begin{figure}[ht]
\resizebox{0.4\textwidth}{!}{\includegraphics{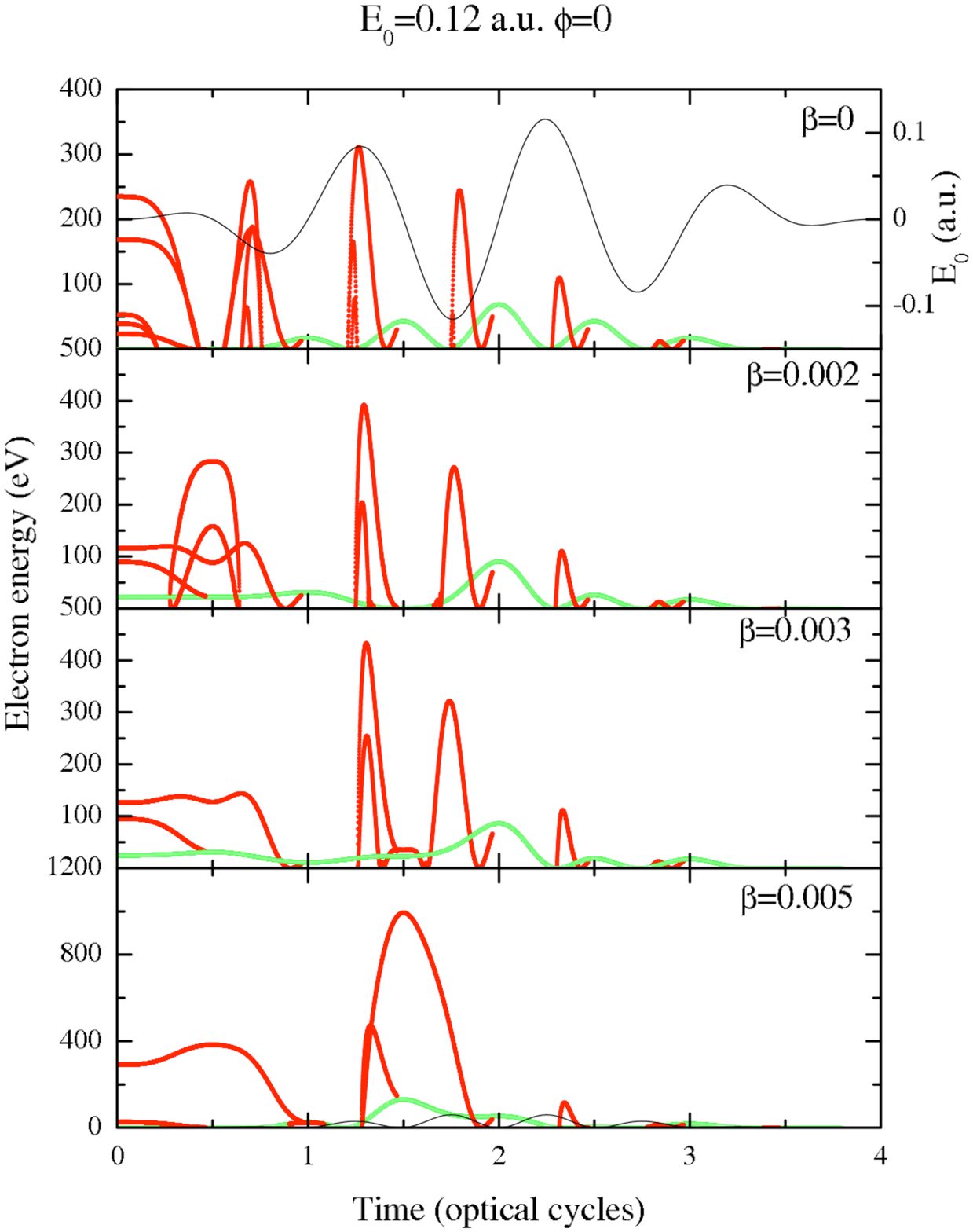}}
\caption{(Color online) Numerical solutions of the Newton equation [Eq. \ref{newton}] plotted in terms of the direct and rescattered electron kinetic energy, $E_{k}^{d}$  and $E_{k}^{r}$, respectively. The laser parameters are $I=5.0544\times10^{14}$ W/cm$^{2}$ ($E_0=0.12$ a.u), $\lambda=800$ nm and $\phi=0$. We employ a few-cycle laser pulse with 4 total cycles (10 fs). Different panels correspond to various values of $\beta$ (see labels). Green filled circles: \textit{direct} electrons; red filled circles: \textit{rescattered} electrons.}
\label{Fig16}
\end{figure}

From the curves for $\beta\neq 0$ we can observe the strong
modifications that the nonhomogeneous character of the laser
electric field produces in the kinetic energy of the laser-ionized
electron.  Furthermore, it is possible to observe how the kinetic
energy is more sensitive to the CEP. If we analyze, for instance,
the top panels of Figs.~12-15 (i.e the homogeneous case) we
conclude that the shape of both the kinetic energy of the direct
and rescattered electron are identical for $\phi=0$ ($\phi=\pi/2$)
and $\phi=\pi$ ($\phi=3\pi/2$). On the other hand, for the
different values of $\beta$ the kinetic energy has a unique shape
for a given value of $\phi$.

The particular features present for $\beta\neq 0$ are related to
the changes in the laser-ionized electron trajectories (for
details see e.g.~\cite{yavuz,ciappi2012,ciappi_opt}). In summary,
the electron trajectories are modified in such a way that now the
electron ionizes at an earlier time and recombines later, and in
this way it spends more time in the continuum acquiring energy
from the laser electric field. Consequently, higher values of the
kinetic energy are attained. This distinct behavior is more
evident for $E_0=0.12$ a.u. and $\beta=0.005$, but it appears to
some degree for all the studied cases.

\begin{figure}[ht]
\resizebox{0.37\textwidth}{!}{\includegraphics{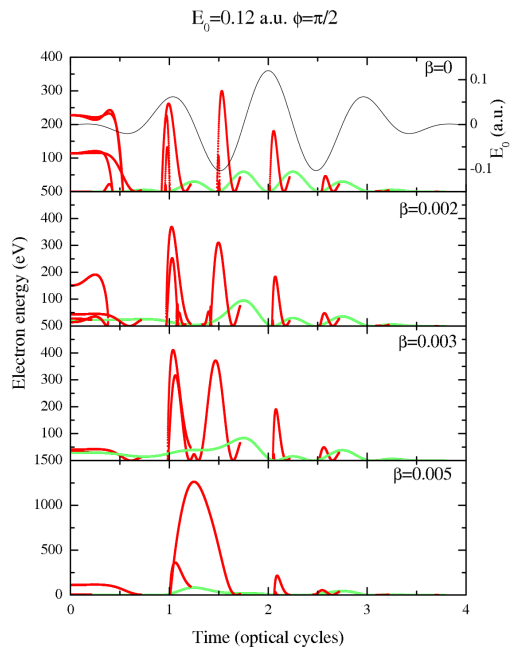}}
\caption{(Color online) Idem Fig. 16 for $\phi=\pi/2$.}
\label{Fig17}
\end{figure}

A similar behavior was observed recently in above threshold
photoemission (ATP) using metal nanotips. According to the model
developed in Ref.~\cite{ropers} the strong localized fields modify
the electron motion in such a way to allow sub-cycle dynamics.

\begin{figure}[ht]
\resizebox{0.37\textwidth}{!}{\includegraphics{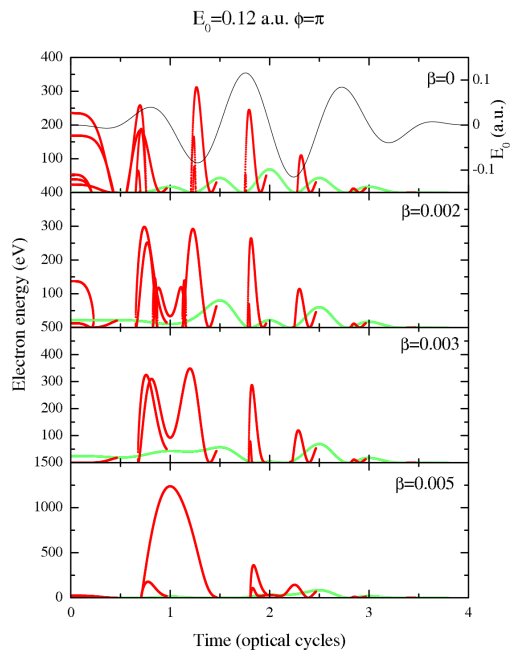}}
\caption{(Color online) Idem Fig. 16 for $\phi=\pi$.}
\label{Fig18}
\end{figure}

In our approach, however, we include the full picture of the ATI
phenomenon, namely both direct and rescattered electrons are
considered (in Ref.~\cite{ropers} only direct electrons are taken
into account) and consequently the characterization of the
dynamics of the photoelectrons is more complex. Nevertheless, the
higher kinetic energy of the rescattered electrons is a clear
consequence of the strong modifications the laser electric field
produces in the region where the electron dynamics takes place, as
in the above mentioned case of ATP.

\begin{figure}[ht]
\resizebox{0.4\textwidth}{!}{\includegraphics{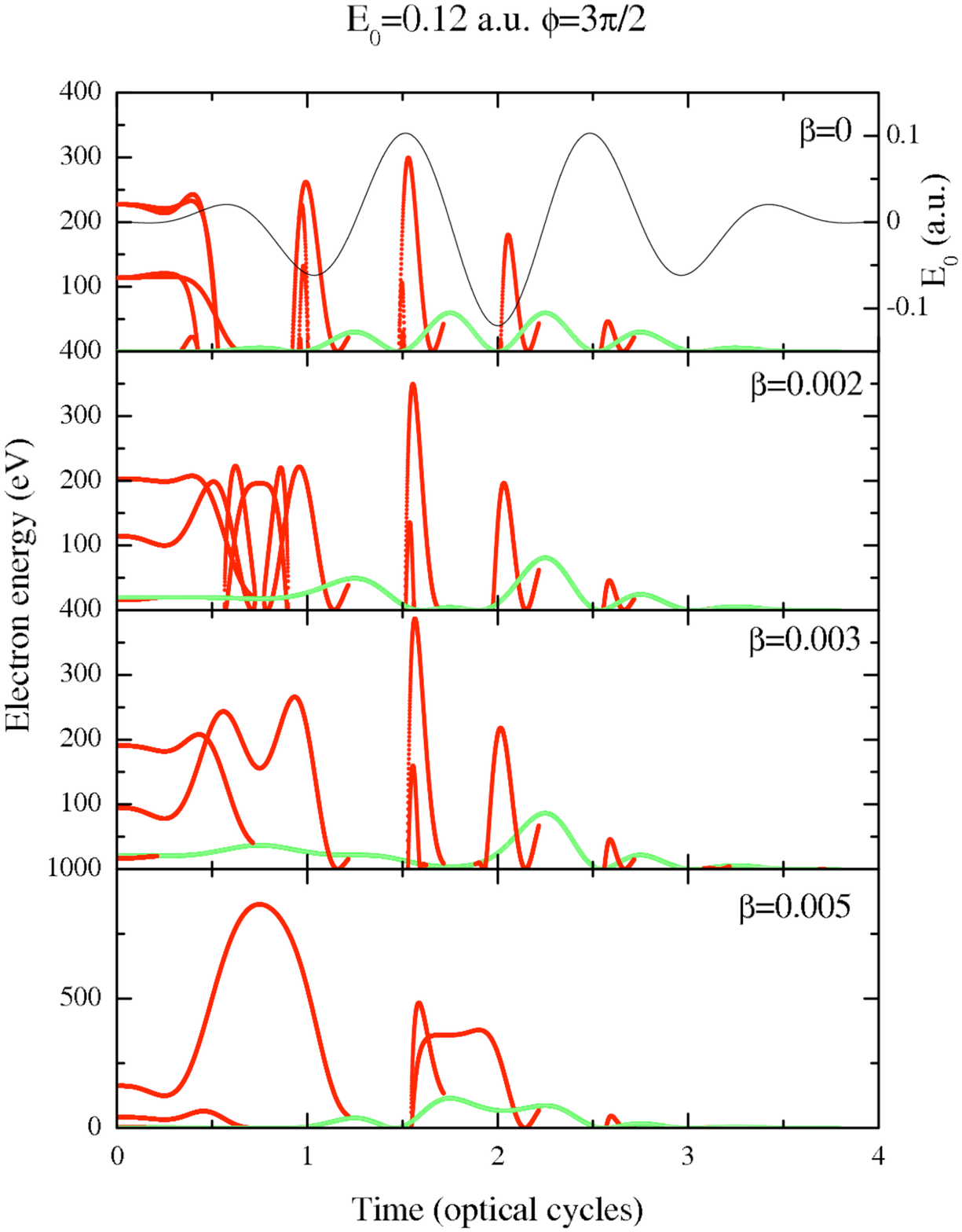}}
\caption{(Color online) Idem Fig. 16 for $\phi=3\pi/2$.}
\label{Fig19}
\end{figure}

\subsection{Multiphoton regime}

For the multiphoton case, we consider a few-cycle laser pulse with
6 complete optical cycles and $E_0=0.05$ a.u.
($I=8.775\times10^{13}$ W/cm$^{2}$) and $\omega=0.25$ a.u.
($\lambda=182.5$ nm). In here, the Keldysh parameter is
$\gamma=5$, indicating the predominance of the multiphoton
process~\cite{arbo1}. We have computed the $P(E)$, two-dimensional
electron distributions and the classical electron energies for all
the set of cases presented in Sec. III.A. In this paper, however,
we just present the most extreme case with an inhomogeneity factor
of $\beta=0.005$ and CEP of $\phi=\pi/2$. These results are
presented in Figs. 20, 21 and 22 for $P(E)$, two-dimensional
electron distributions and the classical electron energies,
respectively. The $P(E)$ exhibits the usual multiphoton
peaks~\cite{Agostini1979,schaferwop1} and the inhomogeneity of the
field does not play any significant role. In the whole range, the
values of the yields have a difference of less than 5\%  and in a
logarithmic scale this is hard to discern.

\begin{figure}[ht]
\resizebox{0.45\textwidth}{!}{\includegraphics{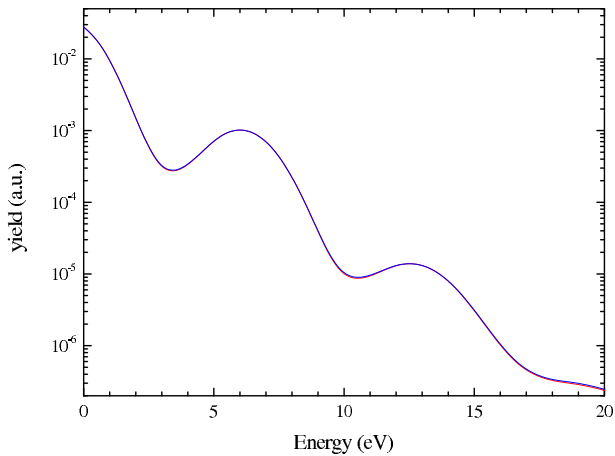}}
\caption{(Color online) Energy-resolved photoelectron spectra $P(E)$ calculated
using the 3D-TDSE for an hydrogen atom ($I_p = -0.5$ a.u.). The laser
parameters are $E_0=0.05$ a.u. ($I=8.775\times10^{13}$ W/cm$^{2}$) and $\omega=0.25$ a.u. ($\lambda=182.5$ nm). We have used
a sin-squared shaped pulse with a total duration of six optical cycles and $\phi=\pi/2$. Red line: homogeneous case ($\beta=0$); blue line: $\beta=0.005$;.}
\label{Fig20}
\end{figure}

The two-dimensional electron distributions are also the same in
terms of shape and magnitude for both homogeneous and inhomogeneous
cases, as shown in Fig.~21. It means the differences introduced by
the spatial nonhomogeneous character are practically
imperceptible. We should note that our calculation is practically
identical to the one presented in~\cite{arbo1}.

The numerical solutions of Eq.~(\ref{newton}) as function of the
kinetic energy (in eV) of the direct and rescattered electron is
depicted in Fig. 22. In here, we could also observe, in support to
our quantum mechanical calculations, that the inhomogeneity of the
field does not change the electron energies in both the direct and
rescattered processes.

In general, we do not find noticeable differences in these
observable quantities for both variations in the CEP and the
strength of the inhomogeneity parameter $\beta$. As a result, we
conclude that in the multiphoton regime the modifications
introduced by the spatial inhomogeneities field do not produce
appreciable modifications in the electron dynamics and
consequently in the measurable quantities. In addition, the
laser-ionized electron, in both the direct and rescattered
processes, has a very small kinetic energy in the multiphoton
regime due to low intensity field. Indeed, in this case the
maximum energy after the rescattering process has taken place is
$\approx 3$ eV. As a result, it is reasonable to have a very small
or almost no differences between the final kinetic energies,  when
a spatial inhomogeneity of small strength  is present.

\begin{figure}[ht]
\resizebox{0.5\textwidth}{!}{\includegraphics[angle=270]{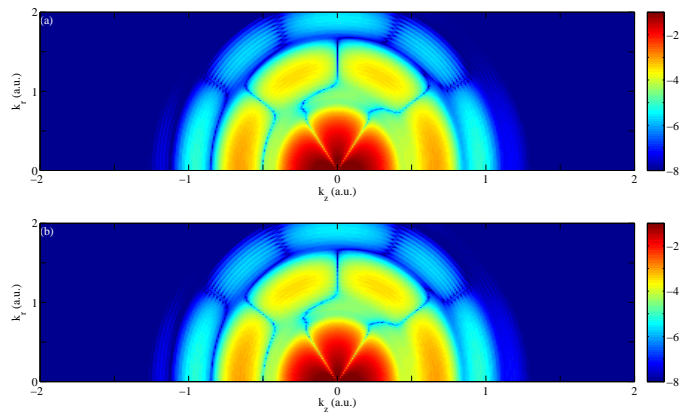}}
\caption{(Color online) Two-dimensional electron momentum
distributions (logarithmic scale) in cylindrical coordinates ($k_z,k_r$)
using the exact 3D-TDSE calculation for an hydrogen atom.
The laser parameters are $E_0=0.05$ a.u. ($I=8.775\times10^{13}$ W/cm$^{2}$), $\omega=0.25$ a.u. ($\lambda=182.5$ nm) and $\phi=\pi/2$. We employ a laser pulse with 6 total cycles. Panel (a) corresponds to the homogeneous case ($\beta=0$) and panel (b) is for $\beta=0.005$.}
\label{Fig21}
\end{figure}

\begin{figure}[ht]
\resizebox{0.45\textwidth}{!}{\includegraphics{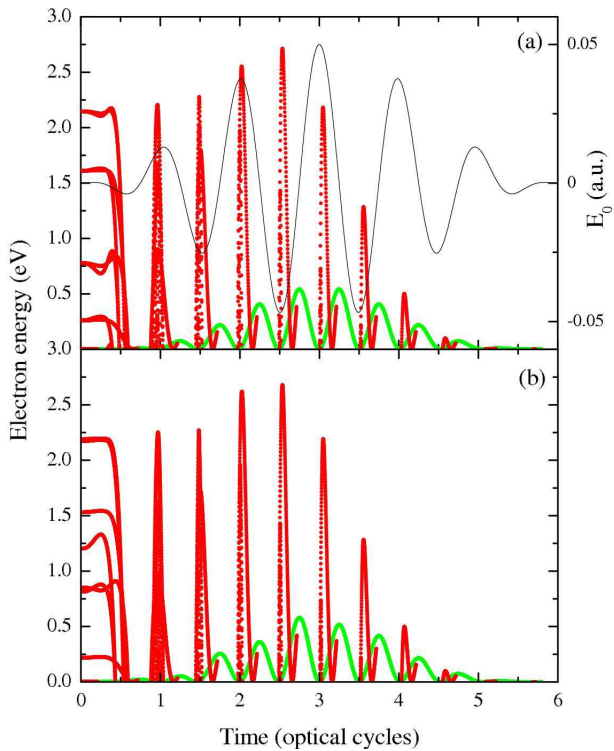}}
\caption{(Color online) Numerical solutions of the Newton equation [Eq.~(\ref{newton})] plotted in terms of the direct and rescattered electron kinetic energy, $E_{k}^{d}$  and $E_{k}^{r}$, respectively. The laser parameters are $E_0=0.05$ a.u. ($I=8.775\times10^{13}$ W/cm$^{2}$), $\omega=0.25$ a.u. ($\lambda=182.5$ nm) and $\phi=\pi/2$. We employ a laser pulse with 6 total cycles. Panel (a) corresponds to the homogeneous case ($\beta=0$) and panel (b) is for $\beta=0.005$. Green filled circles: \textit{direct} electrons; red filled circles: \textit{rescattered} electrons.}
\label{Fig22}
\end{figure}

\section{Conclusions and Outlook}

We have extended our previous studies of ATI produced by
nonhomogeneous fields using the three dimensional solutions of the
TDSE. We have modified the 3D-TDSE to model the ATI phenomenon
driven by spatial nonhomogeneous fields by including an additional
term in the laser-atom coupling. In the tunneling regime we
predict an extension in the cutoff position and an increase of the
yield of the energy-resolved photoelectron spectra in certain
regions. In addition, both the photoelectron spectra and the
two-dimensional electron momentum distributions appear to be more
sensitive to the carrier envelope phase of the laser electric
field. This feature indicates that the ATI produced by spatial
inhomogeneous field could be a good candidate for few-cycle laser
pulse characterization. Furthermore, our predictions pave the way
for the production of high-energy photoelectrons, reaching the keV
regime, using plasmon enhanced fields. In the multiphoton regime,
we show that both the $P(E)$ and the two-dimensional electron
distributions are hardly affected by the spatial nonhomogeneities
of the laser electric field. Our quantum mechanical calculations
are supported by the classical simulations. In particular, the
$P(E)$ characteristics are reasonably well reproduced by
simulations based on classical physics.

\section*{Acknowledgments}

We acknowledge the financial support of the MINCIN/MINECO projects
(FIS2008-00784 TOQATA) (M. F. C. and M.L.); ERC Advanced Grant
QUAGATUA, Alexander von Humboldt Foundation (M. L.); J. A. P.-H.
and L. R. acknowledge support from Spanish MINECO through the
Consolider Program SAUUL (CSD2007-00013) and research project
FIS2009-09522, from Junta de Castilla y Le\'on through the Program
for Groups of Excellence (GR27) and from the ERC Seventh Framework
Programme (LASERLAB-EUROPE, grant agreement n 228334); L. R.
acknowledges the Junta de Castilla y Le\'on through the project
CLP421A12-1.; this research has been partially supported by
Fundaci\'o Privada Cellex.


\end{document}